\documentclass[a4,11pt]{article}

\usepackage{amsmath}
\usepackage{amsfonts}
\usepackage{amssymb}
\usepackage{latexsym}
\usepackage{multicol}
\usepackage{color}
\usepackage[utf8]{inputenc}
\usepackage{graphicx}
\usepackage{dsfont}
\usepackage[T1]{fontenc}
\usepackage{natbib}
\usepackage{placeins}
\usepackage[table]{xcolor}

\def \beg{\begin{eqnarray}}
\def \en{\end{eqnarray}}
\def \be*{\begin{eqnarray*}}
\def\e*{\end{eqnarray*}}

\def\bit{\begin{itemize}}
\def \eit{\end{itemize}}

\newtheorem{rmk}{Remark}[section]

\begin{document}

\begin{center}
{\Large
	{\sc
	 Likelihood-Free Parallel Tempering.
	}
}
\bigskip

Meïli Baragatti$^{1,2,*}$, Agnès Grimaud$^{2}$, Denys Pommeret$^{2}$

\medskip
{\it
 $^1$ Ipsogen SA, Luminy Biotech Entreprises, Case 923, Campus de Luminy, 13288 Marseille Cedex 9, France.\\
 $^2$ Institut de Mathématiques de Luminy (IML), CNRS Marseille, case 907, Campus de Luminy, 13288 Marseille Cedex 9, France.\\
 $^*$ baragatt@iml.univ-mrs.fr, baragattimeili@hotmail.com.
}
\end{center}

\vspace{1cm}

\begin{center}
{\sc March 2012}
\end{center}

\bigskip
\noindent

\begin{abstract}
Approximate Bayesian Computational (ABC) methods, or likelihood-free methods, have appeared in the past fifteen years as useful methods to perform Bayesian analysis when the likelihood is analytically or computationally intractable. Several ABC methods have been proposed:  MCMC methods have been developed by \cite{Marjoram2003} and by \cite{Bortot2007} for instance, and sequential methods have been proposed among others by \cite{Sisson2007}, \cite{Beaumont2009} and \cite{DelMoral2009}. Recently,  sequential ABC methods have appeared  as an alternative to ABC-PMC methods \citep[see for instance][]{McKinley2009,Sisson2007}. In this paper a new algorithm combining population-based MCMC methods with ABC requirements is proposed, using an analogy with the parallel tempering algorithm \citep{Geyer1991}. Performance  is compared with existing ABC algorithms on simulations and on a real example.
\end{abstract}
{\it Keywords}: Approximated Bayesian Computational, likelihood-free, intractable likelihood, parallel tempering, population-based, Monte Carlo Markov chain.

\newpage

\section{Introduction}
     The principle of Approximate Bayesian Computation (ABC) algorithms is to evaluate the posterior density when the likelihood function is intractable. More precisely, let
    $x \in \mathbb{D} \subseteq \mathds{R}^n$ be an observed vector, $\theta \in \mathds{R}^d$ a parameter of interest, and $\pi(.)$ the prior distribution of $\theta$. The aim of an ABC algorithm is to evaluate the following posterior density:
    \begin{displaymath}
	    \pi(\theta \mid x) \propto f(x \mid \theta)\pi(\theta).
    \end{displaymath}
    The likelihood $f(x \mid \theta)$ is supposed to be analytically or computationally intractable. Hence the first requirement is that its calculation is not necessary. The second requirement is the relative ease of simulation from the model, given a parameter $\theta$.
    These requirements are fulfilled in several areas in which ABC methods have been used, including genetics \citep{Pritchard99,Leuenberger2010}, extreme values theory \citep{Bortot2007}, Gibbs random fields \citep{Grelaud2009} or epidemiology \citep{Tanaka2006,Sisson2007}. \\

    The first ABC algorithm has been proposed by \cite{Tavare97}, and is simply a rejection algorithm. It consists in generating jointly $\theta' \sim \pi$ and $z' \sim f(.|\theta')$. The generated $\theta'$ is accepted if the equality $z'=x$ holds, and is then denoted by $\theta_1$. This process is repeated until $n$ generated $\theta'$ have been accepted.
    The accepted $\{\theta_1, \theta_2, \ldots, \theta_n\}$ form an i.i.d.~sample from the target posterior $\pi(. \mid x)$, that is why this algorithm is often called the exact ABC algorithm.
     It has been extended by \cite{Pritchard99} in order to be applied in continuous state spaces. The idea is to accept a simulated $z'$ if it is sufficiently close to the observed $x$. The condition $z'=x$ is replaced by a distance $\rho(z',x)<\varepsilon$, where  $\rho(.,.)$ is a one dimensional function and $\varepsilon$ is a tolerance level. In case of large datasets $z'$ and $x$, it can be computationally demanding to calculate $\rho(z',x)$. Hence a second refinement is to compare summary statistics instead of the complete datasets, and the condition becomes $\rho(S(z'),S(x))<\varepsilon$, with $S$ a statistic which can be multi-dimensional. The algorithm proposed by \cite{Pritchard99} includes these two refinements and is considered as the standard of the ABC algorithms. This algorithm gives an i.i.d.~sample from the following joint posterior distribution:
     \begin{equation}
       \pi_{\varepsilon}(z,\theta \mid x) \propto \pi(\theta)f(z \mid \theta)\mathds{1}_{\{\rho(S(z),S(x))<\varepsilon\}}(z).
     \end{equation}
     It is usually the marginal in $\theta$ which is of interest; that is $\pi_{\varepsilon}(\theta \mid x) \propto \int \pi_{\varepsilon}(z,\theta \mid x) dz.$
     If $S$ provides a good summary of the data and if $\varepsilon$ is sufficiently small, then $\pi_{\varepsilon}(\theta \mid x)$ can be considered as a reasonable approximation of the target posterior $\pi(\theta \mid x)$. Of course, the quality of the obtained approximation depends on the calibration of the algorithm by choosing $\varepsilon$, $\rho(.,.)$ and $S$.

     The acceptance rates of the standard ABC algorithm can be very low,
      hence several improvements have been proposed, using MCMC or sequential schemes.

     \subsection{ABC-MCMC schemes}
     		Algorithms have been proposed by \cite{Marjoram2003}, \cite{Bortot2007} and \cite{Ratmann2007}, the reference method being the one of \cite{Marjoram2003} who suggested the use of a Markov kernel $q(.|.)$ to propose a new parameter from a previous accepted one. The validity of this algorithm called ABC-MCMC is ensured, as the generated Markov chain $(\theta^{(t)},z^{(t)})$ is ergodic, and its invariant density is the target $\pi_{\varepsilon}(z,\theta \mid x)$. The acceptance rates of this algorithm are higher than those obtained with a standard ABC algorithm, but two drawbacks are observed:
		\begin{itemize}
		      \item[$\bullet$]
			In case of small $\varepsilon$ the sampler can encounter difficulties to move in the parameter space. Indeed, the acceptance rate is proportional to the probability of simulating a $z'$ such that $\rho(S(z'),S(x))<\varepsilon$, which is itself proportional to the likelihood \citep[see][]{Sisson2007}. This probability is relatively low if $\varepsilon$ is {small}, and in practice when the ABC-MCMC enters an area of relatively low probability, the acceptance probabilities to move elsewhere are quite {small} and the sampler encounters difficulties to move from this part of the space. As a consequence, not only the sampler sticks in the distribution tails, but also these tails are not well visited.
		      \item[$\bullet$] The samples obtained are dependent, and often highly correlated. This is inherent in many methods based on MCMC, and not necessarily a problem, but if one needs non-correlated particles, sometimes thinning of the chain should be done.\\
		\end{itemize}

\begin{figure*}
			\begin{center}
				\includegraphics[width=0.85\textwidth]{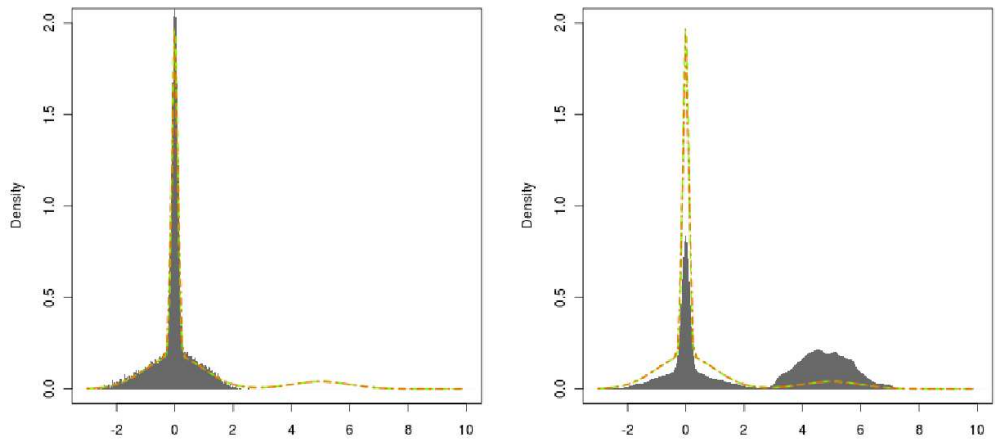}
				\caption{ABC-MCMC for the modified toy example: samples obtained by two runs ($20.10^6$ iterations with $5.10^6$ iterations of burn-in) for tolerance levels of 0.025. The orange dotted curves stand for the exact posterior distribution, the green dotted curves for the target approximation, and densities of obtained samples are represented in gray.}\label{fig:ABCMCMC}
			\end{center}
			\end{figure*}

\begin{figure*}
			 \begin{center}
				  \includegraphics[width=0.5\textwidth]{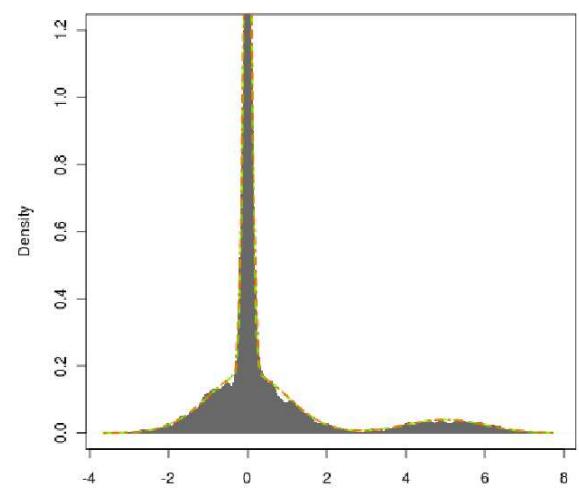}
				  \caption{Modified Toy Example: population associated with a tolerance level of 0.025 (5000 samples) obtained with a standard ABC. The orange dotted curve stands for the exact posterior distributions, the green dotted curve for the target approximation, and the weighted population are represented in gray. In this example the acceptance rate is of 3\%.
				  }\label{fig:ABCstandard}
			  \end{center}
			 \end{figure*}

      \noindent \textit{Modified toy example and ABC-MCMC}\\
		      To illustrate the behavior of the ABC-MCMC in distribution tails, we use a modified toy example inspired from the example studied in \cite{Sisson2007}, \cite{Beaumont2009} and \cite{DelMoral2009}. The prior for $\theta$ is uniform $\mathcal U(-10, 10)$, the model is a mixture of Gaussian distributions given by $x \mid \theta \sim 0.45 \mathcal{N}(\theta,1) + 0.45 \mathcal{N}(\theta,1/100) + 0.1 \mathcal{N}(\theta-5,1)$ (1 and $1/100$ denoting the variances), and the posterior distribution associated with an observation $x=0$ is the mixture:
			\begin{eqnarray*}		
\theta \mid x=0  &\sim&  \\
&&\!\!\!\!\!\!\!\!\!\!\!\!\!\!\!\!\!\!\!\!\!\!\!\!\!\!\Big(0.45 \mathcal{N}(0,1) + 0.45 \mathcal{N}(0,\frac{1}{100}) + 0.1 \mathcal{N}(5,1)\Big)\mathds{1}_{[-10,10]}(\theta).
		\end{eqnarray*}
			Using $S(x)=x$ and $\rho(x,z)=|z-x|$, the approximation of this posterior  $\pi(\theta \mid x=0)$  can be written as:
			\begin{small}
			\begin{eqnarray*}
			 \pi_\varepsilon(\theta \mid x=0) & \propto &\\
 && \!\!\!\!\!\!\!\!\!\!\!\!\!\!\!\!\!\!\!\!\!\!\!\!\!\! 0.45 \Big[ \Phi(\varepsilon-\theta) + \Phi(10(\varepsilon-\theta))- \Phi(-\varepsilon-\theta) - \Phi(10(-\varepsilon-\theta))\Big]
\\
&&\!\!\!\!\!\!\!\!\!\!\!\!\!\!\!\!\!\!\!\!\!\!\!\!\!\! + 0.1 \Big[\Phi(\varepsilon-\theta+5) - \Phi(-\varepsilon-\theta+5)\Big],
			\end{eqnarray*}	
			\end{small}		
where $\Phi$ denotes the distribution function of a $\mathcal N(0,1)$.
			Using a tolerance level $\varepsilon = 0.025$, the approximated posterior is indistinguishable from the exact posterior density. As in \cite{Sisson2007} and \cite{Beaumont2009}, we used a $\mathcal N(\theta^{(t-1)}, 0.15^2)$ as the proposal $q(. \mid \theta^{(t-1)})$.
			The algorithm was run on $20.10^6$ iterations with a burn-in of $5.10^6$, and it appears that for some runs the chain did not succeed to visit the small mode around 5, while for others it stayed stuck in this local mode and does not visit well the global mode around 0. Figure \ref{fig:ABCMCMC} shows the results of two runs. It is clear that these results are not satisfactory. As a comparison, the population obtained by a standard ABC is given in Figure \ref{fig:ABCstandard}. It is quite good, hence the issue observed in the distribution tail is specific to MCMC, it is not due to the ABC approximation of the likelihood.

     	\subsection{Contribution of the paper}
      	Until now ABC methods using MCMC schemes were not completely satisfactory. Hence sequential algorithms have been proposed by different authors:  \cite{Sisson2007}, \cite{Beaumont2009} (similar algorithms have been proposed by \cite{Sisson2009} and \cite{Toni2009}), and more recently \cite{DelMoral2009} (a similar algorithm has been proposed by \cite{Drovandi2011}). The ABC-PMC of \cite{Beaumont2009} is simple to use and to implement, and it gives non-biased results from the approximate posterior $\pi_{\varepsilon}(\theta \mid x)$. A more recent algorithm is the ABC-SMC of \cite{DelMoral2009}: both tolerance levels and transition kernels are automatically calibrated at each iteration, and several datasets are generated given a proposed parameter. These sequential ABC schemes are now very popular. They compare favorably to MCMC schemes, but they have not been compared to MCMC schemes using a population of chains.\\
	In this paper we are interested in developing such a method, by doing an analogy with the Parallel Tempering algorithm used in classical MCMC methods \citep{Geyer1991,GeyerThompson}. In particular, we want to know if using a population-based approach the drawback of ABC-MCMC schemes concerning distribution tails can be avoided. Moreover, we would like to compare the behavior of this approach with  ABC-PMC and ABC-SMC algorithms, especially in distribution tails.\\

	\noindent The paper is organized as follows.  In Section 2 we introduce the new population-based ABC-MCMC scheme, that we call ABC-PT algorithm.
	In Section 3 this algorithm is illustrated on the modified toy example and on real data, and comparisons with the ABC-MCMC, ABC-PMC and ABC-SMC algorithms are done. Finally, the method and the results are discussed in Section~4.

\section{The ABC-PT algorithm}
	\subsection{Standard Parallel Tempering (PT)}
		When the goal is to sample from a multi-modal target density $\pi$, classical MCMC methods like Metropolis-Hastings algorithm  \citep{Metropolis1953,Hastings1970}  or Gibbs sampler \citep{GelfandSmith} for instance, are often trapped into local modes from where they cannot escape in reasonable time. To avoid this problem, the principle of PT is to introduce $N$ temperatures and to run in parallel $N$ associated MCMC chains with target distributions being tempered distributions of the target $\pi$. 
		The first chain targets $\pi$, and since the tempered distributions become flatter as the temperature increases, the chains at high temperatures can move easily between modes and explore the whole parameter space. Each iteration of the PT algorithm is decomposed into two types of moves: local moves via classical MCMC algorithms to update the different chains, and global moves allowing swaps between two chains. The use of these swaps enables new modes to be propagated through the different chains, thereby improving mixing. The first chain associated with the target distribution will then be able to escape from local modes.
	
	\subsection{Analogy with the PT in an ABC framework}
		The analogy with the PT is the following: a population of $N$ chains is used, chains of higher orders should be able to move easily in the parameter space, and chains of lower orders should give precise approximations of the target posterior. These last chains can have some difficulties to exit local modes, so exchange moves between chains should be defined so that the chains of lower orders can visit all the parameter space, and hence can exit low probability areas. \\
		To achieve this in an ABC framework, chains of higher orders are associated with high tolerance levels, while chains of lower orders are associated with low tolerance levels. A sequence of tolerance levels is then defined as  $\boldsymbol{\varepsilon}=(\varepsilon_1,\varepsilon_2,\ldots,\varepsilon_N)$, with $\varepsilon_1<\varepsilon_2<\ldots<\varepsilon_N$. The $i${th} chain is associated with $\varepsilon_i$. In  particular the first chain is associated with $\varepsilon_1$ the tolerance level required for the approximation of the posterior of interest, and the chain associated with the higher tolerance level $\epsilon_N$ should be able to exit areas of low posterior probabilities.

		The set of parameters  associated to the $N$ chains is written $\boldsymbol{\theta}=(\theta_1,\theta_2,\ldots,\theta_N)$ with $\theta_i \in \Theta \subseteq \mathds{R}^d$, and $\boldsymbol{z}=(z_1,z_2,\ldots, z_N)$ denotes the associated simulated datasets. The state associated to the $i${th} chain  is $(z_i,\theta_i)$, for $i \in \{1, \ldots,N\}$.
	
		\subsubsection{Local moves}
			Each chain is locally updated using an iteration of an ABC-MCMC algorithm, with associated tolerance level and transition kernel. In order to improve the exploration of the parameter space by higher order chains, the transition kernels associated with these chains should enable large moves. As the tolerance levels associated with these chains are high, the probabilities to accept the proposed moves can remain reasonably high. On the opposite, transition kernels associated with lower order chains should propose smaller moves, to have more chance to accept the proposed moves. We then propose to use tempered transition kernels, using a sequence of temperatures $T_1=1<T_2<\ldots<T_N$ (such tempered transitions kernels have been used by \cite{Ratmann2007}). In particular the $i${th} chain is associated with $\varepsilon_i$, $T_i$ and a symmetric transition kernel $q_i(. \mid .)$. For instance $q_i(. \mid .)$ can be a Gaussian transition kernel tempered by $T_i$. This $i${th} chain is associated with the following joint distribution:
			 	  \begin{equation*}
			 	    \pi_{\varepsilon_i}(z_i,\theta_i \mid x) \propto \pi(\theta_i)f(z_i \mid
			 \theta_i)\mathds{1}_{\{\rho(S(z_i),S(x))<\varepsilon_i\}}(z_i).
			 	  \end{equation*}
			 	  We also introduce:
			 	  \begin{equation*}
			 	   \pi^*_{\varepsilon}(\boldsymbol{z},\boldsymbol{\theta} \mid x) = \prod_{i=1}^N \pi_{\varepsilon_i}(z_i,\theta_i
			 \mid x).
			 	  \end{equation*}
				
		\subsubsection{Exchange moves}

		  To perform an exchange move between two chains, a pair of chains is chosen uniformly. The orders of these  chains are denoted by $i$ and $j$ ($i<j$), and the states proposed by the exchange move are written $\boldsymbol{\theta}'$ and $\boldsymbol{z}'$. We denote by $q(\boldsymbol{z}',\boldsymbol{\theta}' \mid \boldsymbol{z},\boldsymbol{\theta})$ the probability of proposing $(\boldsymbol{z}',\boldsymbol{\theta}')$ from $(\boldsymbol{z},\boldsymbol{\theta})$. 
		  We propose to exchange the state $(z_i,\theta_i)$ of chain $i$ with the state $(z_{j},\theta_{j})$ of chain $j$. We  then have:
		  \begin{eqnarray*}
	 	  \boldsymbol{\theta}'&=&(\theta_1,\ldots,\theta_j,\ldots,\theta_i,\ldots,\theta_N),  \\ \boldsymbol{z}'&=&(z_1,\ldots,z_j,\ldots,z_i,\ldots,z_N).
		  \end{eqnarray*}
		  The acceptance probability to ensure reversibility of the move will be the following (using the classical detailed balance condition):
		  \begin{eqnarray}
	   \alpha &= & 1 \wedge \frac{\pi^*_{\boldsymbol{\varepsilon}}(\boldsymbol{z}',\boldsymbol{\theta}' \mid x)q(\boldsymbol{z},\boldsymbol{\theta} \mid \boldsymbol{z}',\boldsymbol{\theta}')}{\pi^*_{\boldsymbol{\varepsilon}}(\boldsymbol{z},\boldsymbol{\theta} \mid x)q(\boldsymbol{z}',\boldsymbol{\theta}' \mid \boldsymbol{z},\boldsymbol{\theta})},
  \nonumber\\
  &=&   1 \wedge \frac{\pi^*_{\boldsymbol{\varepsilon}}(\boldsymbol{z}',\boldsymbol{\theta}' \mid x)}{\pi^*_{\boldsymbol{\varepsilon}}(\boldsymbol{z},\boldsymbol{\theta} \mid x)}.
\label{acceptPTABCwrong}		
\end{eqnarray}
		  Indeed, as the pair of chains is chosen uniformly among all the possible pairs, it is clear that $q(\boldsymbol{z},\boldsymbol{\theta} \mid \boldsymbol{z}',\boldsymbol{\theta}')=q(\boldsymbol{z}',\boldsymbol{\theta}' \mid \boldsymbol{z},\boldsymbol{\theta})$.
		  In addition we have:
		  \begin{eqnarray}
		    \nonumber \frac{\pi^*_{\boldsymbol{\varepsilon}}(\boldsymbol{z}',\boldsymbol{\theta}' \mid x)}{\pi^*_{\boldsymbol{\varepsilon}}(\boldsymbol{z},\boldsymbol{\theta} \mid x)} && \\ && \!\!\!\!\!\!\!\!\!\!\!\!\!\!\!\!\!\!\!\!\!\!=\frac{\pi_{\varepsilon_i}(z_i',\theta_i' \mid x)\pi_{\varepsilon_j}(z_j',\theta_j' \mid x)}{\pi_{\varepsilon_i}(z_i,\theta_i \mid x)\pi_{\varepsilon_j}(z_j,\theta_j \mid x)},
\nonumber\\
		    &&
\!\!\!\!\!\!\!\!\!\!\!\!\!\!\!\!\!\!\!\!\!\!= \frac{f(z_i' \mid \theta_i')\mathds{1}_{\{\rho(S(z_i'),S(x))<\varepsilon_i\}}(z_i')}{f(z_i \mid \theta_i)\mathds{1}_{\{\rho(S(z_i),S(x))<\varepsilon_i\}}(z_i)}
\nonumber \\
&&
\!\!\!\!\!\!\!\!\!\!\!\!\!\!\!\!\!\! \times \frac{f(z_j' \mid \theta_j')\mathds{1}_{\{\rho(S(z_j'),S(x))<\varepsilon_j\}}(z_j')}{f(z_j \mid \theta_j)\mathds{1}_{\{\rho(S(z_j),S(x))<\varepsilon_j\}}(z_j)},
\nonumber
\\
		    &&
\!\!\!\!\!\!\!\!\!\!\!\! \!\!\!\!\!\!\!\!\!\!=\frac{\mathds{1}_{\{\rho(S(z_j),S(x))<\varepsilon_i\}}(z_j) \mathds{1}_{\{\rho(S(z_i),S(x))<\varepsilon_j\}}{(z_i)}}{\mathds{1}_{\{\rho(S(z_i),S(x))<\varepsilon_i\}}(z_i)
\mathds{1}_{\{\rho(S(z_j),S(x))<\varepsilon_j\}}(z_j)}.
\label{ratiopietoile}		
\end{eqnarray}
		  By construction $\epsilon_i<\epsilon_j$, and \\$\mathds{1}_{\{\rho(S(z_i),S(x))<\varepsilon_i\}}(z_i) \mathds{1}_{\{\rho(S(z_j),S(x))<\varepsilon_j\}}(z_j)=1$. \\Therefore (\ref{ratiopietoile}) simply reduces to $\mathds{1}_{\{\rho(S(z_j),S(x))<\varepsilon_i\}}(z_j)$, and the exchange moves is accepted if $\{\rho(S(z_j),S(x))<\varepsilon_i\}$. 	
		 This exchange move satisfies ABC requirements as there is no need to compute any likelihood. We can note that this acceptance rate depends only on  the tolerance level $\varepsilon_i$.

    \subsection{ABC-PT algorithm}\label{algoABCPT}
	  Define a sequence of tolerance levels $\varepsilon_1<\varepsilon_2<\ldots<\varepsilon_N$, and a sequence of temperatures $T_1=1<T_2<\ldots<T_N$.
	  \vspace{20pt}\hrule\vspace{3pt}
	  \noindent \textbf{ABC-PT} \par\nobreak
	  \vspace{3pt}\hrule\vspace{6pt}
	  \noindent \texttt{To obtain a Markov chain of length $n$:}
	  \begin{enumerate}
	  \item \texttt{$t=1$. Use the standard ABC algorithm to obtain $N$ realizations: \\For $i$ in $\{1,\ldots,N\}$,\\
	      generate $(\theta_i^{(0)},z_i^{(0)})$ from $\pi_{\varepsilon_i}(z,\theta \mid x)$.\\
	      Set $\boldsymbol{\theta}^{(0)}=(\theta_1^{(0)},\theta_2^{(0)},\ldots,\theta_N^{(0)})$ and $\boldsymbol{z}^{(0)}=(z_1^{(0)},z_2^{(0)},\ldots,z_N^{(0)})$.}\\

	  \item \texttt{\noindent For $t$ in $\{2,\ldots,n\}$:}
	      \begin{enumerate}
	       \item \texttt{LOCAL MOVES ABC-MCMC: For $i$ in $\{1,\ldots,N\}$:}
		  \texttt{\begin{enumerate}
		  \item Generate $\theta_i'$ from $q_i(. \mid \theta_i^{(t-1)})$ depending on $T_i$.
		  \item Generate $z_i'$ given $\theta_i'$, from  $f(.\mid \theta_i')$.
		  \end{enumerate}}
		  \hspace{0.3cm} \texttt{Set $(\theta_i^{(t)},z_i^{(t)})=(\theta_i',z_i')$ with probability}
		  \begin{displaymath}
		  \alpha = min\Bigg\{1,\frac{\pi(\theta_i')q_i\Big(\theta_i^{(t-1)} \mid \theta_i'\Big)}{\pi(\theta_i^{(t-1)})q_i\Big(\theta_i' \mid \theta_i^{(t-1)}\Big)}\Bigg\}\mathds{1}_{\{\rho(S(z_i'),S(x))<\varepsilon_i\}}(z_i').
		  \end{displaymath}
		  \hspace{0.3cm} \texttt{Else set $(\theta_i^{(t)},z_i^{(t)})=(\theta_i^{(t-1)},z_i^{(t-1)})$.}\\

	      \item \texttt{EXCHANGE MOVES:\\  $N$ exchange moves are proposed: \\ $N$ pairs of chains are chosen uniformly in all possible pairs with replacement. \\ For each of them:
	      Denote by $i$ and $j$ ($i<j$) the orders of a chosen pair of chains.
\\ Exchange $(z_i,\theta_i)$ with $(z_j,\theta_{j})$ \\ if  $\rho(S(z_j),S(x))<\varepsilon_i$.}
	      \end{enumerate}
	   \end{enumerate}
	  \vspace{3pt}\hrule\vspace{20pt}

	  At each iteration $N$ exchange moves are proposed. Indeed, it appears from our experience that only one proposition is not sufficient to enable a good mixing of the first chain of interest. Moreover increasing the number of exchange moves improves the mixing with the first chain of interest and allows to obtain less correlated samples.

   \subsection{Improving acceptance rates in ABC-PT}\label{ring}
	In order to increase the acceptance rate of the exchange moves in the ABC-PT algorithm, we  consider a $K$-partition of the tolerance levels space, for a fixed integer $K$.  In this way we obtain $K$ disjoint subsets  $E_1, \cdots, E_K$,  that we shall call {\it rings}. A chain of order $i$ is associated to a ring $E_j$ if $\rho(S(z_i),S(x))\in E_j$. In the exchange move part of the ABC-PT algorithm,  two chains  are then chosen such that they are associated to the same ring.
	Thus, the second part of the algorithm is modified as follows:

	\vspace{5pt}
\hrule\vspace{10pt}
		      (b) \texttt{EXCHANGE MOVES WITH RINGS:  \\$N$ exchange moves are proposed: \\ For each exchange move a ring with at least two associated chains is chosen randomly and two associated chains are chosen uniformly in this ring. \\ For each of them:
		      Denote by $i$ and $j$ ($i<j$) the orders of a chosen pair of chains.
	\\ Exchange $(z_i,\theta_i)$ with $(z_j,\theta_{j})$  if $\rho(S(z_j),S(x))<\varepsilon_i$.}
	\vspace{5pt}
\hrule
	\vspace{5pt}

	In practice, taking into account these rings, if an exchange between two selected chains $i$ and $j$ is proposed, then the acceptance probability  $\mathds{1}_{\{\rho(S(z_j),S(x))<\varepsilon_i\}}(z_j)$  is generally greater than in the case where the two chains are chosen uniformly (without rings).
Note that this idea of partitioning can be related to the construction of  rings of energy of the states space proposed in \cite{KouZhouWong}.

\section{Illustrations on modified toy example}
    \subsection{ABC-PT}\label{ModifToyEx-ABCPT}
	We use the modified toy example presented in the Introduction to study the behavior of the ABC-PT algorithm, especially in distribution tails and in local posterior modes.
	We take $N=15$ chains, and imitating  the ABC-MCMC algorithm, the tempered transition kernels used are $q_i(.|\theta_i^{(t-1)}) \sim \mathcal N(\theta_i^{(t-1)}, 0.15^2 T_i)$, $i=1,\ldots,N$.  Concerning the sequence of temperatures, we choose $T_1 =1$ and $T_N = 4$. The other temperatures can be chosen by evenly spacing them on a logarithmic scale, by evenly spacing their inverses, or by evenly spacing their inverses geometrically \citep[see for instance][]{KouZhouWong,NagataWatanabe,Neal1996}. In this example we decide to space them on a logarithmic scale.
	Concerning the tolerance levels, once $\varepsilon_{1}$ and $\varepsilon_N$ are chosen, it gives satisfactory results to choose the others by using a sequence of proportionality constants \citep[approach similar to the one of ][]{Beaumont2009} or by regularly spacing them on a logarithmic scale (which is a special case of a sequence of proportionality constants). In this example they are chosen evenly spaced on a logarithmic scale between $\varepsilon_1= 0.025$ and $\varepsilon_N =2$. We checked that the highest order chain associated with $T_N = 4$ and $\varepsilon_N =2$ easily explores the parameter space. The algorithm is run for 600 000 iterations with a burn-in period of 150 000 iterations.

\begin{figure*}
	    \begin{center}
		    \includegraphics[width=1\textwidth]{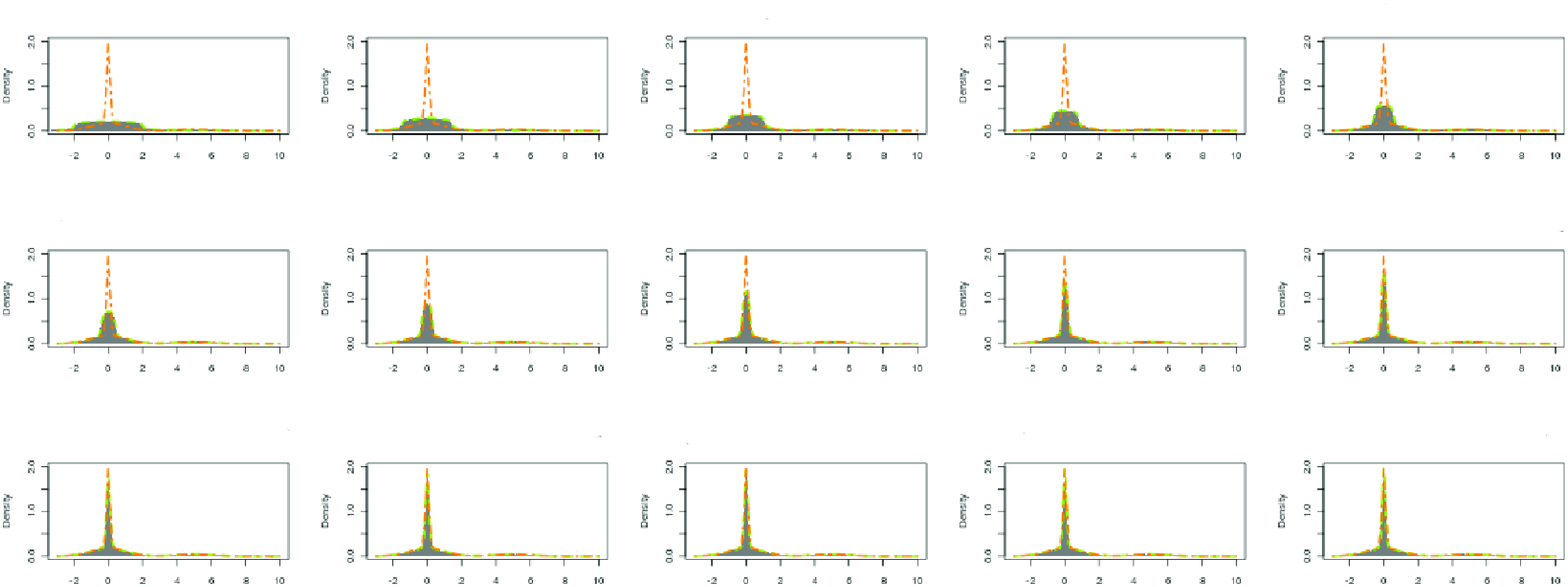}
		    \caption{ABC-PT for the modified toy example: approximations obtained by each of the $N$ chains, the associated sequence of tolerance levels is $\varepsilon=(2, 1.4625, 1.0694, 0.7820, 0.5719, 0.4182, 0.3058, 0.2236, 0.1635, 0.1196, 0.0874, 0.0639, 0.0468, 0.0342, 0.025)$, for 600 000 iterations with a burn-in of 150 000. The orange dotted curves stand for the exact posterior distribution, the green dotted curves for the target approximation, and densities of obtained samples are represented in gray.}\label{fig:ToyExampleBis5_PTABC}
	    \end{center}
	    \end{figure*}

\begin{figure*}
 	    \begin{center}
 		    \includegraphics[width=0.45\textwidth]{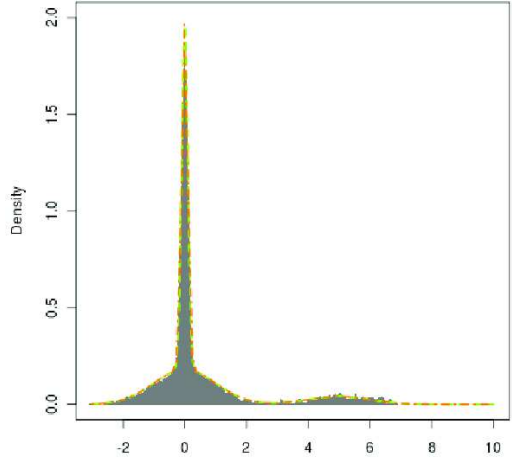}
 		    \caption{ABC-PT for the modified toy example: approximations obtained by the chain associated with tolerance level of 0.025 (600 000 iterations with a burn-in of 150 000). The orange dotted curves stand for the exact posterior distribution, the green dotted curves for the target approximation, and densities of obtained samples are represented in gray.}\label{fig:ToyExampleBis5_PTABC2zoom}
 	    \end{center}
 	    \end{figure*}

	Figures \ref{fig:ToyExampleBis5_PTABC} and \ref{fig:ToyExampleBis5_PTABC2zoom} show the obtained approximations $\pi_{\epsilon_i}(.|x)$ for the $N$ chains in parallel. They are satisfactory, the two modes being well visited.
	Table \ref{tab:ABCPT_actu} presents the acceptance rates and the numbers of accepted exchange moves for six of the 15 chains:  on average, 2.05 moves are accepted per iteration (out of 15).

	    \begin{table*}
	    \begin{center}
	    \begin{tabular}{|c|c|c|c|c|c|c|}
	    \hline
	    $\epsilon$ & 2 & 0.78 & 0.31 & 0.12 & 0.047 & 0.025 \\
	    \hline
	    Local acceptance rates  & 65.8\% & 45.6\% & 26.0\% & 12.5\% & 5.4\% & 3.0\%\\
	    Accepted exchange moves  & 106 711 & 171 678 & 191 311 & 187 557 & 156 676 & 107 330\\
	    \hline
	    \end{tabular}
	    \caption{Modified toy example: acceptance rates for local moves and number of accepted exchange moves, for chains 15, 12, 9, 6, 3 and 1 of the ABC-PT.}\label{tab:ABCPT_actu}	    \end{center}
	    \end{table*}

\FloatBarrier	

\begin{figure*}
 	  \begin{center}
 		  \includegraphics[width=0.75\textwidth]{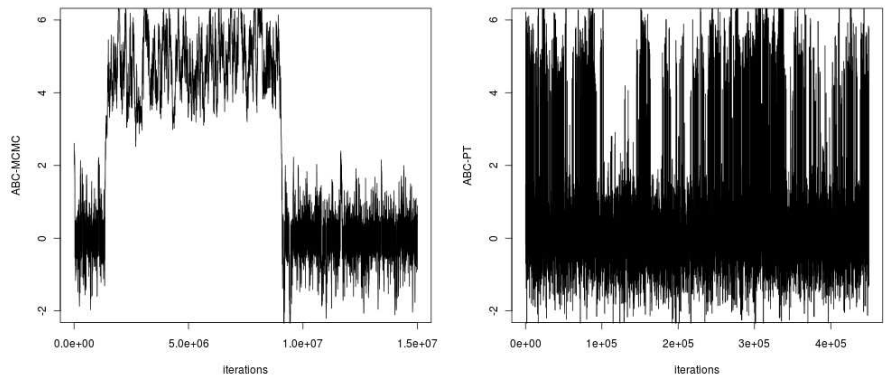}
 		  \caption{Modified toy example: On the left the trace of a chain obtained by ABC-MCMC (example corresponding to the right part of Figure \ref{fig:ABCMCMC}), on the right the trace of a first chain obtained by ABC-PT.}\label{ToyExModif5_traceABCMCMC-ABCPT}
	  \end{center}
 	  \end{figure*}

	Compared with the approximations obtained with the ABC-MCMC algorithm, the results obtained with ABC-PT are greatly improved in terms of exploration of the parameter space. Indeed, the first chain of interest succeeds to visit the small mode around 5, and it is not trapped in this local mode. Moreover, the results of different runs appeared very similar indicating that the results of ABC-PT are reproducible.
Table \ref{ToyExampleBis5Corr} shows that there is also an improvement in terms of autocorrelations. 
	In Figure \ref{ToyExModif5_traceABCMCMC-ABCPT} the traces of the two algorithms are given, and it is observed that the mixing of the chain is improved by ABC-PT compared to ABC-MCMC.
	
	 \begin{table*}
	  \begin{center}
	  \begin{small}
	  \begin{tabular}{|c|c|c|c|c|c|}
	  \hline
	  & thinning & Nb of samples & Order 1 & Order 10 & Order 20\\
	  \hline
	   & none & $15.10^6$ & 1 & 1 & 1\\
	  ABC-MCMC & 100 & $15.10^4$ & 0.999 & 0.994 & 0.990\\
	   & 1000 & $15.10^3$ & 0.994 & 0.968 & 0.951\\
	  \hline
	  & none & $45.10^4$ & 0.842 & 0.367 & 0.284\\
	  ABC-PT (chain 1) & 10 & $45.10^3$ & 0.361 & 0.198 & 0.142\\
	   & 50 & $9.10^3$ & 0.215 & 0.093 & 0.082\\
	  \hline
	  \end{tabular}
	  \caption{Modified toy example: autocorrelations of order 1, 10 and 20 of the chain obtained by ABC-MCMC, and of the first chain obtained by ABC-PT (with or without thinning the chains).}\label{ToyExampleBis5Corr}
	  \end{small}
	  \end{center}
	  \end{table*}
	
\FloatBarrier

      \subsection{Comparison with ABC-PMC}
\begin{figure*}
	      \begin{center}
		      \includegraphics[width=1\textwidth]{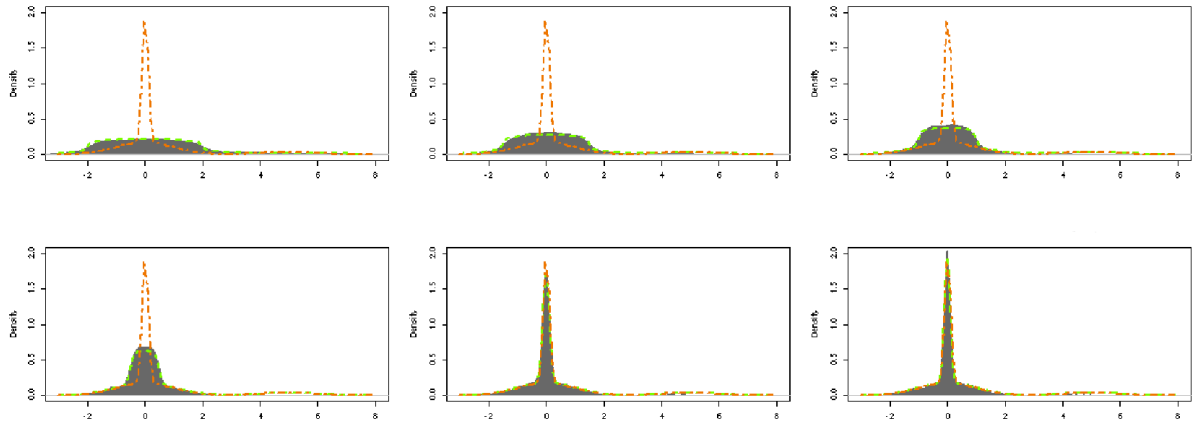}
		      \caption{ABC-PMC for the modified toy example: the six populations are associated with tolerance levels of 2, 1.5, 1, 0.5, 0.1 and 0.025 (5000 samples). The orange dotted curves stand for the exact posterior distribution, the green dotted curves for the target approximation, and the weighted populations are represented in gray.}\label{fig:ABC-PMC}
	      \end{center}
	      \end{figure*}

\begin{figure*}
	      \begin{center}
		      \includegraphics[width=0.9\textwidth]{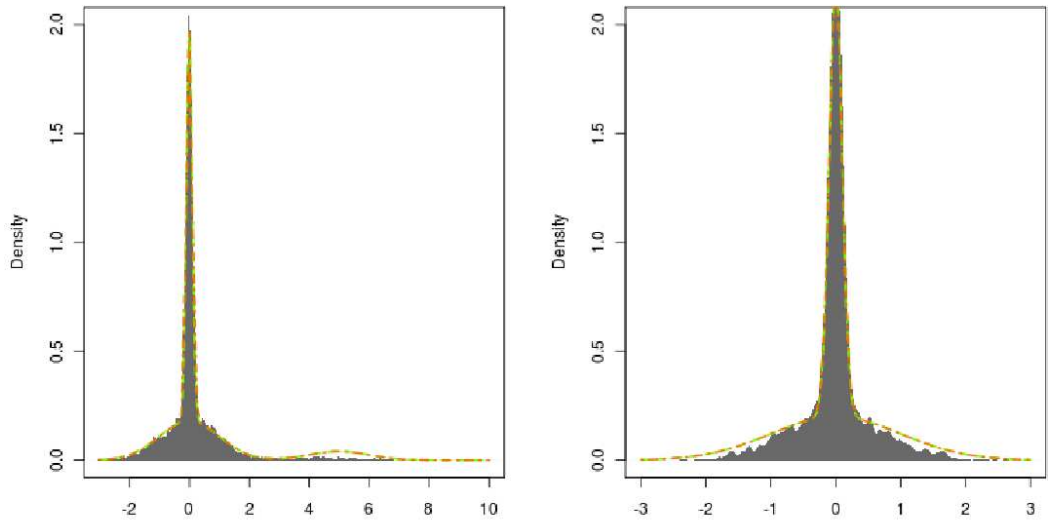}
		      \caption{ABC-PMC zoom: populations associated with a tolerance level of 0.025 (5000 samples), for the modified toy example on the left, and for the standard toy example on the right. The orange dotted curves stand for the exact posterior distributions, the green dotted curves for the target approximation, and the weighted populations are represented in gray.
		      }\label{fig:ABC-PMCzoom}
	      \end{center}
	      \end{figure*}

	  The ABC-PMC of \cite{Beaumont2009} is run on the modified toy example. The decreasing sequence of tolerance levels used is the same than in \cite{Beaumont2009}, that is $(2, 1.5, 1, 0.5, 0.1, 0.025)$. Six weighted populations of size 5000 are generated, and represented in Figure \ref{fig:ABC-PMC}. The results are much better than those obtained by the ABC-MCMC. In particular, the sampler is not stuck in distribution tails. However it seems that as the tolerance level decreases, the tails are under-covered and in particular the small mode around 5, see the left of Figure \ref{fig:ABC-PMCzoom}. That is also observed for the standard toy example (without the small mode around 5), see the right of Figure \ref{fig:ABC-PMCzoom}.

	\begin{rmk}
	The ABC-PMC algorithm has been run with the same sequence of tolerance levels than those used for the ABC-PT: 15 populations were generated, hence it took more time than a run in which only a sequence of six tolerance levels is used. It appeared that the 15{th} weighted population associated with $\varepsilon_1=0.025$ did not cover better the local mode around 5. Moreover the diversity of this population was decreased, as its Effective Sample Size was of 3858.295, compared to 4164.55 for the 6{th} population when six populations were generated. Note that like the ABC-PT,  the results of the ABC-PMC  were reproducible.
	\end{rmk}

      \subsection{Comparison with ABC-SMC}
\begin{figure*}
 	  \begin{center}
 		  \includegraphics[width=0.95\textwidth]{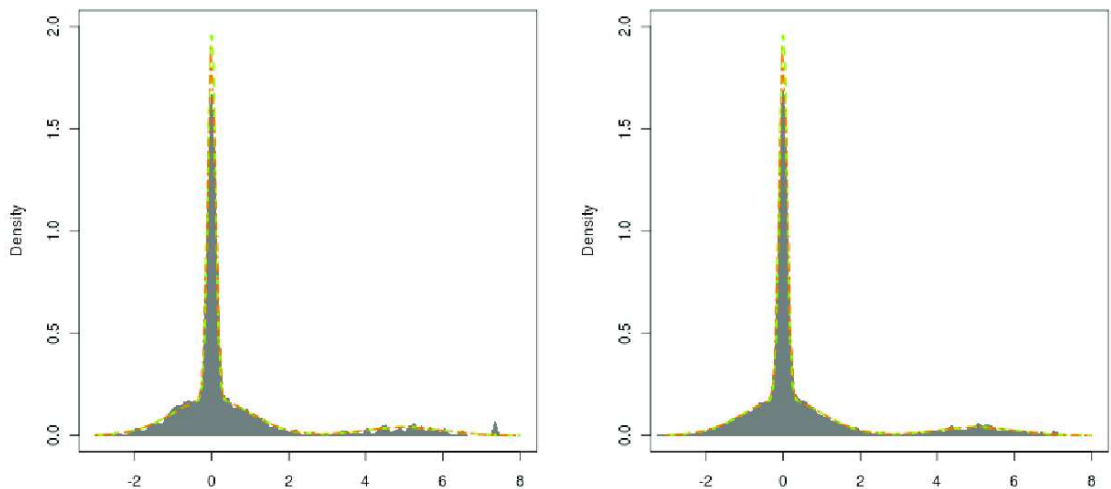}
 		  \caption{ABC-SMC for the modified toy example, for a tolerance level of 0.025: on the left the results for $M=1$ and 15 000 particles,   on the right the results for $M=10$ and 20 000 particles. The orange dotted curves stand for the exact posterior distribution, the green dotted curves for the target approximation, and densities of obtained samples are represented in gray.}\label{ToyExampleBis5_SMC}
	  \end{center}
 	  \end{figure*}

	We also compare the results of the ABC-PT with those obtained with the ABC-SMC \citep{DelMoral2009}. 
	We take $\epsilon=0.025$ and $\alpha= 0.95$. To run ABC-SMC and ABC-PMC with similar CPU times, we first take $M=1$ and 15 000 particles, see the left of Figure \ref{ToyExampleBis5_SMC}. The Effective Sample Size is of 11242. We also run the ABC-SMC for $M=10$ and 20 000 particles, see the right of \ref{ToyExampleBis5_SMC}. The Effective Sample Size is of 14666.98. Generally, it appears that the results of the ABC-SMC are reproducible and similar to those obtained by the ABC-PT. However, we can note two differences: using the ABC-SMC the global mode around 0 is slightly under-covered, and the ABC-SMC has the advantage to give non-correlated samples.

\FloatBarrier

	  \begin{rmk}
	  In an ABC framework, all sequential methods require simulating a number of pseudo-observations proportional to the number of particles. But their computational costs differ concerning the calculation of some importance weights: this cost is quadratic in the number of particles for the algorithms of \cite{Beaumont2009} (ABC-PMC), \cite{Sisson2007} and \cite{Toni2009}, while it is linear for the adaptive ABC-SMC of \cite{DelMoral2009} \citep[see also][]{BeskosCrisanJasra}. Comparison with the ABC-MCMC or the ABC-PT is not easy. Indeed, MCMC methods give unweighted correlated samples, hence the use of the effective sample size is not relevant. To compare the sequential and MCMC approaches, comprehensive convergence studies of the MCMC methods in an ABC framework should be achieved, taking into account the number $N$ of chains, the temperature spacings \citep[see][]{Atchade2010}, as well as the autocorrelations.
	  \end{rmk}

      \subsection{Comparison with $N$ independent parallel chains}
\begin{figure*}
 	  \begin{center}
 		  \includegraphics[width=1\textwidth]{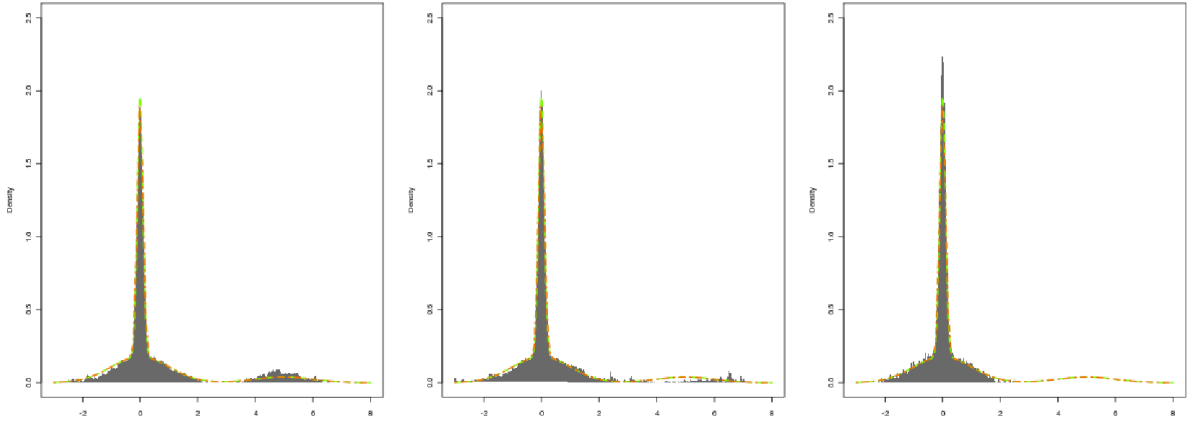}
 		  \caption{Modified toy example: three runs of 15 independent parallel chains with tolerance levels of 0.025 (600 000 iterations with a burn-in of 150 000 for each chain, hence we obtain 6 750 000 post-burn-in samples).}\label{ToyExampleBis5_resNinde}
 	  \end{center}
 	  \end{figure*}

	Finally we compare the results of the ABC-PT with those obtained with $N$ independent parallel chains. The 15 independent chains are locally updated by an ABC-MCMC algorithm during 600 000 iterations with a burn-in period of 150 000 iterations, with $\varepsilon = 0.025$ and $\mathcal N(\theta^{(t-1)}, 0.15^2)$ as the proposal $q(. \mid \theta^{(t-1)})$. Figure \ref{ToyExampleBis5_resNinde} shows on three runs that the results obtained for the $N$ independent chains are not reproducible: the run represented on the left enables to detect the small mode around 5 (which is a little bit over-covered), the run on the middle detect the small mode which is quite under-covered, while the run on the right did not detect it at all (none of the 15 chains succeeded to visit this mode). The traces of each of the 15 chains are similar to those of the ABC-MCMC algorithm, see the left of Figure \ref{ToyExModif5_traceABCMCMC-ABCPT}. On the opposite, it appears that thanks to the exchange moves, the ABC-PT always succeeds to visit the small mode, and to escape it.
	
      \subsection{ABC-PT using rings}\label{rings}

\begin{figure*}
 	    \begin{center}
 		    \includegraphics[width=0.5\textwidth]{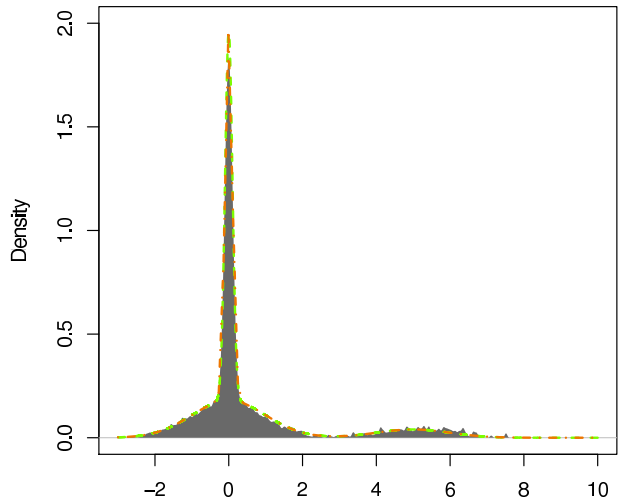}
		    \caption{Approximations obtained by the first chain after an ABC-PT algorithm with $N=15$ chains, with 3 rings of tolerance levels  (600 000 iterations with a burn-in of 150 000). The orange dotted curves stand for the exact posterior distribution, the green dotted curves for the target approximation, and densities of obtained samples are represented in gray.}\label{fig:anneau}
	    \end{center}
	    \end{figure*}

	We illustrate the approach with rings on the modified toy example. As previously we use $N=15$, $T_N=4$, $\epsilon_1=0.025$ and $\epsilon_{N}=2$. We choose $K=3$ rings containing the same number of tolerance levels. We get $E_1=[0,0.103], E_2=]0.103,0.495]$, and $E_3=]0.495,2]$.
	The algorithm is run for 600 000 iterations with a burn-in period of 150 000 iterations.
	The results are given in Figure \ref{fig:anneau}. Compared to the ABC-PT without rings, we observe that for an average of about $+13\%$ of CPU times, an increase of about $+83\%$ of accepted exchange moves is obtained.  All the $N(N-1)/2$ exchange rates between chains increase with the use of rings, and especially the rates between chains with high temperatures. Matrices of accepted exchange rates obtained on two runs of ABC-PT (with or without rings) are given in Appendix (Tables  \ref{matrixABCPTring} and \ref{matrixABCPT}). An improvement in terms of autocorrelation is also observed as shown in Table \ref{tab:anneau}.

	  \begin{table*}
	  \begin{center}
	  \begin{small}
	  \begin{tabular}{|c|c|c|c|c|c|}
	  \hline
	  & thinning & Nb of samples & Order 1 & Order 10 & Order 20\\
	  \hline
	  & none & $45.10^4$ & 0.622 & 0.264 & 0.238\\
	  ABC-PT (chain 1) & 10 & $45.10^3$ & 0.268 & 0.148 & 0.111\\
	  & 50 & $9.10^3$ & 0.193 & 0.070 & 0.048\\
	  \hline
	  \end{tabular}
	  \caption{Modified toy example: autocorrelations of order 1, 10 and 20 of the first chain obtained by ABC-PT with 3 rings of tolerance levels (with or without thinning the chains).}\label{tab:anneau}
	  \end{small}
	  \end{center}
	  \end{table*}

\FloatBarrier

    \section{Illustration on tuberculosis data}
	We consider the real example of tuberculosis transmission studied by \cite{Tanaka2006}, \cite{Sisson2007} and \cite{Blum2010}.
	The data come from a study of a tuberculosis epidemic in San Francisco during 1991 and 1992 \citep{Small94}. They consist of DNA fingerprint at the IS6110 marker for 473 isolates, which are grouped in 326 distinct genotypes represented as follows:
	$$30^1\; 23^1\; 15^1\; 10^1 \; 8^1 \; 5^2 \; 4^4\; 3^{13} \; 2^{20} \; 1^{282},$$
	where $n^k$ indicates that $k$ clusters of size $n$ were observed.

	A model of disease transmission and marker mutation is defined, which is an extension of a birth and death process: a birth corresponds to a new infection and a death corresponds to a host dead or healed. A system of stochastic differential equations is used, and is governed by three rates: $\alpha$ the birth rate per case per year, $\delta$ the death rate per case per year and $\theta$ the mutation rate per case per year (see Appendix \ref{formuleTuberculose}). However interest parameters for biologists are the transmission rate $\alpha-\delta$, the doubling time $\log(2)/(\alpha - \delta)$ and the reproductive value $\alpha/\delta$, results will then be given for these parameters in the following.

	The likelihood of this model can not be written explicitly. However it is easy (although long) to simulate datasets from this model \citep[see][for details]{Tanaka2006}. This example is then particularly suited for the use of ABC methods. We use the same simulation process than in \cite{Tanaka2006}.

	To compare simulated and observed data, two summary statistics considered as relevant by the biologists are used: the number of distinct genotypes in the studied sample $g$, and the gene diversity $H$ defined by $H=1-\sum_{i=1}^g(\frac{n_i}{n})^2$, where $n_i$ is the size of cluster with genotype $i$ and $n$ is the size of the sample. The values $g_{obs}$ and $H_{obs}$ denote the two statistics obtained on the observed data. A simulated sample of size $n$ associated to $g$ and $H$ is considered close to the observed data if $\frac{1}{n}|g_{obs}-g|+ |H_{obs}-H| < \varepsilon$, where $\varepsilon$ is a suitable tolerance level.

	Prior distributions for the parameters $\alpha$ and $\delta$ are uniforms on the interval $[0, 5]$ with $\alpha >\delta$.
	Concerning the mutation rate $\theta$, many studies have been conducted to estimate it \citep[see][for references]{Tanaka2006}, which enable us to take the following informative prior: $\theta \sim \mathcal N(0.198, 0.06735^2)$ truncated on the left at $0$.

	Concerning the ABC-PT algorithm, the vector of parameters is denoted by $\phi =(\alpha, \delta, \theta)$, and $\phi_i^{(t)}$ is the vector of the chain $i$ at time $t$, for $i=1, ..., N$.
	As in \cite{Tanaka2006}, the transition kernel for the first chain $q_1(.|\phi_1^{(t-1)})$ corresponds to a $\mathcal N(\phi_1^{(t-1)}, \Sigma)$, with:
	$$\Sigma =
	\left(
	\begin{array}{ccc}
	0.5^2 & 0.225  & 0  \\
	0.225 &  0.5^2 &   0\\
	0 & 0  & 0.015^2
	\end{array}
	\right).$$
	For the other chains, $q_i(. \mid \phi_i^{(t-1)})$ corresponds to a $\mathcal{N}(\phi_i^{(t-1)},\Sigma_i)$, with $\Sigma_i=\Sigma^{1/T_i}$.
	We took  $N=7$ chains, associated with temperatures between $T_1 =1$ and $T_N = 2$ evenly spaced on a logarithmic scale.
	We used a Quad-Core Xeon E5320 1.86GHz processor with 8GB of RAM and we chose a minimum tolerance level of 0.01 to have reasonable acceptance rates for generated chains \citep[to be compared with the tolerance level of 0.0025 taken by ][]{Tanaka2006}.
	The sequence of tolerance levels was evenly spaced on a logarithmic scale as follows: $\varepsilon_1=0.01, \varepsilon_2=0.03, \cdots,  \varepsilon_7=0.1$.

	\paragraph{Results}~~\\

	The algorithm was run for 20 000 iterations with a burn-in period of 2 000 iterations, it took about 4.5 days.
	The first chain was locally updated in 3.8 \% of the iterations, and accepted on average 0.25 exchange move per iteration (5095 total moves). Concerning the set of $N=7$ chains, on average 3.07 exchange moves were accepted per iteration (among 7). Table \ref{tab:TuberculoseABCPTech} gives the local acceptance rates for each chain, and proportions of accepted exchange moves for each pair of chain (among the proposed exchange moves).
	      \begin{table*}
	      \begin{center}
	      \begin{tabular}{|c|ccccccc|}
	      \hline
	       & chain 1 & chain 2 & chain 3  & chain 4  & chain 5  & chain 6  & chain 7\\
	      \hline
	      chain 1 & 3.8 & 22.6 & 17.0 & 13.2 & 10.7 & 7.6 & 5.6\\
	      chain 2 & . & 13.9 & 76.1 & 58.8 & 44.5 & 34.0 & 25.2\\
	      chain 3 & . & . & 16.2 & 77.0 & 58.9 & 43.9 & 33.6\\
	      chain 4 & . & . & . & 17.9 & 77.3 & 56.3 & 45.0 \\
	      chain 5 & . & . & . & . & 18.9 & 74.3 & 59.9\\
	      chain 6 & . & . & . & . & . & 19.9 & 78.3\\
	      chain 7 & . & . & . & . & . & . & 20.4\\
	      \hline
	      \end{tabular}
	      \caption{ABC-PT for the tuberculosis example: the diagonal gives local acceptance rates, and outside the diagonal are given the proportions of accepted exchange moves.}\label{tab:TuberculoseABCPTech}
	      \end{center}
	      \end{table*}

\begin{figure*}
	    \begin{center}
		    \includegraphics[width=1\textwidth]{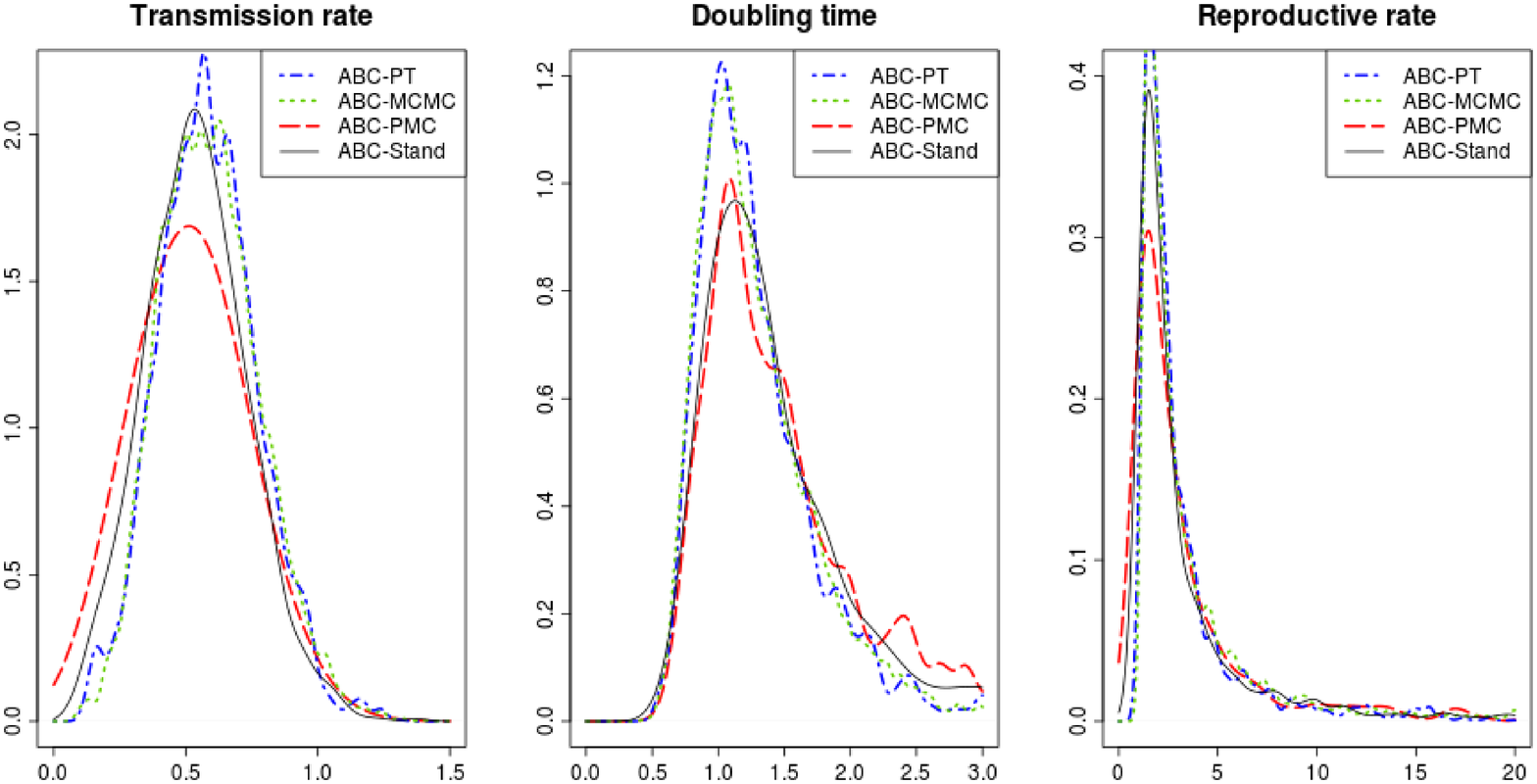}
		    \caption{Posterior densities of the transmission rate, the doubling time, and the reproductive rate, obtained by the algorithms ABC-PT, ABC-MCMC, ABC-MCMC and standard ABC.}\label{graphCompTuber}
	    \end{center}
	    \end{figure*}

	Table \ref{tab:TuberculoseABCPT} gives the posterior estimates of the interest parameters, obtained from the first chain. These estimates are of the same orders than those obtained by \cite{Tanaka2006}, \cite{Sisson2007} and \cite{Blum2010}. The mean transmission rate is around 0.60, describing the rate of increase
	of the number of cases in the population. The doubling time gives the same information, as it is just a transformation of the transmission rate. This parameter is used by biologists, as it can be easily interpreted (period of time required to double the percentages of cases in the population). Here it is estimated around 1.35 years. The credibility interval of the reproductive value (number of new infections generated by one case) is quite large, as we have positive posterior probabilities for few very small values of $\delta$, corresponding to high values of this rate. The mean is then unstable. However, most of the posterior values are around the median (see Figure \ref{graphCompTuber}), which should then be preferred compared to the mean.
	      \begin{scriptsize}
	      \begin{table*}
	      \hspace*{-2cm}
	      \begin{tabular}{|c|c|c|c|c|c|c|c|c|}
	      \hline
	      & \multicolumn{3}{c|}{ABC-PT results} & \multicolumn{3}{c|}{\cite{Tanaka2006} results} & \multicolumn{2}{c|}{\cite{Blum2010} results}\\
	      \hline
	       & Mean & Median & 95\% CI & Mean & Median & 95\% CI & Posterior Mode & 95\% CI\\
	      \hline
	      Transmission rate & 0.59 & 0.58 & (0.29,0.92) & 0.69 & 0.68 & (0.38-1.08) & 0.56 & (0.16-0.95)\\
	      Doubling time & 1.35 & 1.20 & (0.76,2.41) & 1.08 & 1.02 & (0.64-1.82) & 1.16 & (0.73-4.35)\\
	      Reproductive rate & 5.98 & 2.29 & (1.20,17.35) & 19.04 & 3.43 & (1.39-79.71) & 4.00 & (2.24-117.45)\\
	      Mutation rate & 0.25 & 0.25 & (0.15,0.35) & & & & & \\
	      \hline
	      \end{tabular}
	      \caption{ABC-PT for the tuberculosis example: posterior estimates of the parameters of interest, on the 18 000 post-burn-in iterations of the first chain: mean, median and 95\% credibility interval.}\label{tab:TuberculoseABCPT}
	      \end{table*}
	      \end{scriptsize}

	We then compared the results of the ABC-PT algorithm with those of the standard ABC, ABC-MCMC and ABC-PMC algorithms. The same minimum tolerance level of 0.01 is used for all the algorithms. Concerning standard ABC, 1000 samples were generated; concerning ABC-MCMC 350 000 iterations were performed with a burn-in period of 50 000 iterations; and concerning ABC-PMC 10 populations of 1000 samples were generated using the following sequence of tolerance levels \citep[from][]{Sisson2007}: $\varepsilon_1 = 1, \varepsilon_{10} = 0.01$ and $\varepsilon_k = 0.5(\varepsilon_{k-1}+\varepsilon_{10}), k=2,\ldots,9$. Note that for the ABC-PMC algorithm we used  weighted covariance matrices as suggested in \cite{Filippi2011} to get optimal kernels.

	Posterior estimates of parameters obtained by these three algorithms were close to those of the ABC-PT, see
	Figure \ref{graphCompTuber}.

\begin{figure*}
 	  \begin{center}
 		  \includegraphics[width=0.75\textwidth]{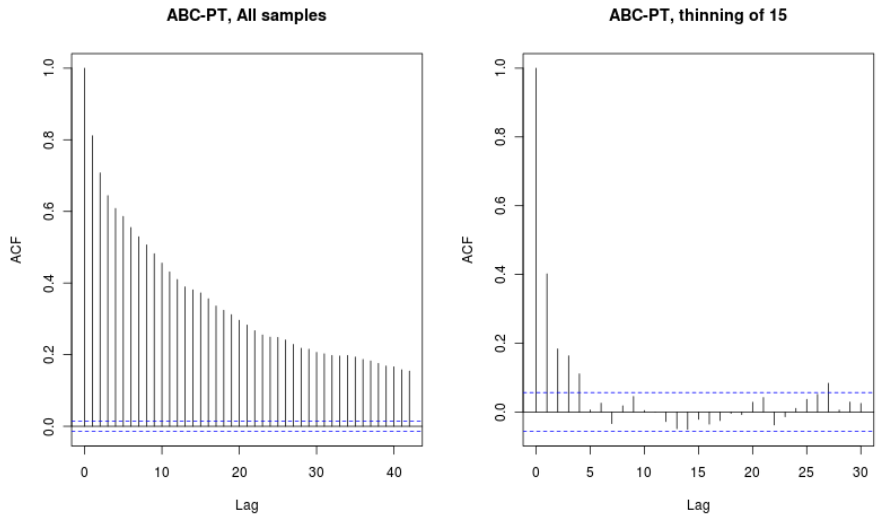}
 		  \caption{For the $\alpha$ parameter, autocorrelation plots of the first chain obtained by ABC-PT: on the left all samples are kept, on the right every 15{th} sample is kept.}\label{autocorPT2}
 	  \end{center}
 	  \end{figure*}

	\begin{itemize}
	\item[$\bullet$] Standard ABC algorithm: an average number of 408 data generation steps were needed to obtain one sample. This algorithm took approximately 6.5 days.
	\item[$\bullet$] ABC-PMC algorithm: 203 data generation steps were necessary on average to obtain one particle and computing time was about 4.5 days.
	\item[$\bullet$] ABC-MCMC algorithm: 9 days were required.  The chain was updated in 3.8 \% of iterations. The obtained simulations are more correlated than those of the first chain of the ABC-PT algorithm. See Table \ref{tuberculoseCorr}, which gives for the $\alpha$ parameter the autocorrelations of the chain obtained by ABC-MCMC, and of the first chain obtained by ABC-PT, with or without thinning. Similar results were obtained for the parameters $\delta$ et $\theta$. Hence an improvement of the ABC-PT over the ABC-MCMC in terms of autocorrelations was observed. If one is interested in having samples not too correlated, it appears from Table \ref{tuberculoseCorr} that in order to have autocorrelations decreasing rapidly and non significant from the fifth order, every 500{th} sample must be kept for the ABC-MCMC algorithm, and every 15{th} sample must be kept for ABC-PT. As a consequence, 583 (350 000/600) and 117 (20 000 $\times$ 7 / 1200) data generation steps are required on average to obtain one realization of the target posterior from the ABC-MCMC and the ABC-PT respectively. Figure \ref{autocorPT2} gives the autocorrelation plot for the ABC-PT when all the samples are kept, and when a thinning of 15 is used.
	\end{itemize}

  \begin{table}[!h]
	  \begin{center}
	  \begin{small}
	  \begin{tabular}{|c|c|c|c|c|c|}
	  \hline
	  & thinning & Nb of samples & Order 1 & Order 10 & Order 20\\
	  \hline
	   & none & 300 000 & 0.991 & 0.919 & 0.850\\
	  ABC-MCMC & 100 & 3 000 & 0.513 & 0.107 & 0.032\\
	   & 500 & 600 & 0.162 & 0.056 & 0.017\\
	  \hline
	  & none & 18 000 & 0.812 & 0.456 & 0.296\\
	  ABC-PT (chain 1) & 10 & 1 800 & 0.473 & 0.007 & -0.040\\
	  & 15 & 1 200 & 0.402 & 0.005 & 0.029\\
	  \hline
	  \end{tabular}
	  \caption{For the $\alpha$ parameter: autocorrelations of order 1, 10 and 20 of the chain obtained by ABC-MCMC, and of the first chain obtained by ABC-PT (with or without thinning the chains).}\label{tuberculoseCorr}
	  \end{small}
	  \end{center}
	  \end{table}
	
	Note that the obtention of correlated samples is part of any MCMC method, and satisfactory estimates can nevertheless be achieved.

\section{Discussion}
    The ABC-PT algorithm developed in this paper is a new likelihood-free method based on the Markov Chain Monte Carlo theory. On the studied examples, results obtained by the algorithm are better than those obtained by the classical ABC-MCMC algorithm, in terms of exploration of the parameter space, and of autocorrelations of the generated chain. The performances of this new algorithm appear equivalent to those of the ABC-PMC and of the ABC-SMC, which are sequential methods. These sequential methods have the advantage to generate non-correlated samples. However, on the modified toy example the ABC-PMC appeared to slightly under-cover posterior distribution tails and the local mode, while the ABC-SMC appeared to slightly under-cover the global mode. The choice between ABC-PT, ABC-PMC and ABC-SMC should then depend on the simulation's objective. We would recommend to use the ABC-PT or the ABC-SMC if we suspect that the posterior of interest is multi-modal, with small modes in low probability areas. 
    Concerning the computational cost of the ABC-PT, a convergence study should be achieved, taking into account the number $N$ of chains, the temperature spacings \citep[see][]{Atchade2010}, as well as the autocorrelations.

    From a theoretical point of view the $N$ chains of the ABC-PT all together can be considered as a Markov chain, and the ABC-PT algorithm generates a Markov chain for $(\boldsymbol{z},\boldsymbol{\theta})$ with stationary distribution $\pi^*_{\boldsymbol{\varepsilon}}(\boldsymbol{z},\boldsymbol{\theta} \mid x)$. The chain obtained for $\varepsilon_1$ is the chain of interest, because it provides samples corresponding to $\pi_{\varepsilon_1}(z_1,\theta_1 \mid x)$ which is the target distribution. The marginal in $\theta_1$ $\pi_{\varepsilon_1}(\theta_1 \mid x)$ can be considered as a good approximation of the posterior of interest $\pi(. \mid x)$ if $\varepsilon_1$ is small enough and $S$ well chosen according to our data.

    From a practical point of view, the ABC-PT requires calibration of the sequence of tolerance levels, and of the sequence of temperatures. Practical guidelines have been given in the modified toy example, see section \ref{ModifToyEx-ABCPT}.
    As seen in section \ref{rings}, it is possible to improve the acceptance rate of the exchange moves by considering a partition of the tolerance levels space to choose the pairs of chains to exchange. The total number of accepted exchange moves was increased by about $83\%$, with only $13\%$ of additional CPU time.

    In this paper the ABC-PT algorithm only uses one type of global move, which is the exchange move. However, it is always possible to define other types of global moves, like those of the Evolutionary Monte Carlo of \cite{LiangWong} for instance. It is also possible to define delayed rejection moves, see for instance \cite{GreenMira} and \cite{JasraStephensHolmes2007b}.
    It would be of particular interest to use the simulations generated by the other chains than the first chain, by developing a method in the same spirit than those of \cite{Beaumont2002} or \cite{BlumFrancois2010}. Moreover, as proposed by a reviewer, we can take advantage of using a schedule of varying  $\varepsilon_1$ for the first chain of interest, to improve its mixing. Finally, we think that using a pre-determined sequence of tolerance levels is not the best way to obtain a pre-determined number of samples with a tolerance level as small as possible, for a given computational time. In the lines of \cite{MullerSansoDeIorio} it would then be interesting to use a non pre-determined sequence of tolerance levels, and hence to have an adaptive algorithm.


\paragraph{{\bf Acknowledgements}} The authors are very grateful to the reviewers and to the Associate Editor for useful comments which enabled to greatly improve the manuscript.

\bibliography{referencesabbrev}

\appendix

\section{Matrices of exchange rates for ABC-PT with and without rings}\label{AnnexeMatEch}
Tables \ref{matrixABCPTring} and \ref{matrixABCPT} compare accepted exchange rates between chains based on the modified toy example, using ABC-PT algorithms with or without rings.  The rate between chains $i$ and $j$ is equal to the number of accepted exchange moves between these chains divided by the total number of iterations. It is observed that the use of rings allows more accepted exchange moves, especially between high tempered chains.

     \begin{table}[!h]
	      \begin{footnotesize}
	      \begin{tabular}{|c|cccccccccccccc|}
	      \hline
	       &  2 &  3  & 4  & 5  & 6  & 7 & 8 & 9 & 10 & 11 & 12 & 13 & 14 & 15 \\
	      \hline
	      chain 1  & 0.16 &  0.12 &  0.09 &  0.06 &  0.04 &  0.03 &
0.02 &  0.01 &  0.01 &  0.01 &  0.01 &  0.00 &
 0.00 &  0.00
\\
	      chain 2 & . &  0.16 &  0.12 &  0.09 &  0.06 &  0.04 &
0.03 & 0.02&  0.01 & 0.01 &  0.01 &  0.01 &
0.00 &  0.00
\\
	      chain 3 & . & . & 0.16 &  0.12 &  0.08 &  0.056 &
0.04 &  0.03 & 0.02 &  0.01 & 0.01 &  0.01 &
0.01 &  0.00
\\
	      chain 4 & . & . & . &  0.16 & 0.11 &  0.077  &
0.05 &  0.04 & 0.03 &  0.02 &  0.01 &  0.01 &
0.01 &  0.01
\\
	      chain 5 & . & . & . & . & 0.15   & 0.10 &
0.07 &  0.05 &  0.04 &  0.03 &  0.02 &  0.02 &
0.01 &  0.01
\\
	      chain 6 & . & . & . & . & . & 0.11  &
0.08 &  0.05 &  0.04 &  0.03 &  0.02 & 0.02 &
0.01 &  0.01
\\
	      chain 7 & . & . & . & . & . & . &
 0.10 & 0.07 &  0.06 &  0.04 &  0.03 &  0.02 &
0.01 &  0.01
\\
 chain 8 & . & . & . & . & . & . &
 & 0.15 & 0.11 &  0.08 &  0.05 &  0.04 &
 0.02 &  0.02
 \\
  chain 9 & . & . & . & . & . & . &
 .&.& 0.23 &  0.16 & 0.11 &  0.07 &
  0.05 &  0.04
  \\
   chain 10 & . & . & . & . & . & . &
  . & . &.& 0.30 &  0.19 &  0.13 &
  0.09 &  0.06
\\
   chain 11 & . & . & . & . & . & . &
  .&.&.&.&  0.22 &  0.15 &
     0.11 &  0.08
   \\
   chain 12 & . & . & . & . & . & . &
  . &.&.&.&. &  0.24 &
     0.18 &  0.14
   \\
    chain 13 & . & . & . & . & . & . & .
    &. &. &. &. &. &
      0.43 &  0.34
    \\	
  chain 14 & . & . & . & . & . & . & .
&. &. &. &. &. &
    . &  0.77
   \\
 \hline
	      \end{tabular}
	      \end{footnotesize}
	      \caption{ABC-PT with 3 rings for the modified toy example: the diagonal gives local acceptance rates, and outside the diagonal are given the proportions of accepted exchange moves.}\label{matrixABCPTring}
	      \end{table}

     \begin{table}[!h]
	      \begin{footnotesize}
	      \begin{tabular}{|c|cccccccccccccc|}
	      \hline
	       &  2 &  3  & 4  & 5  & 6  & 7 & 8 & 9 & 10 & 11 & 12 & 13 & 14 & 15 \\
	      \hline
	      chain 1  &  0.10 &  0.08 &  0.06 &  0.04 &  0.03 &  0.02
 &
 0.02 & 0.01&  0.01&  0.01& 0.00 &  0.00
 &
 0.00 &  0.00
\\
	      chain 2 & . &  0.10 &  0.08 &  0.06 &  0.04 &  0.03 &
 0.02 &  0.02 & 0.01 &  0.01 & 0.01 & 0.00
 &
0.00 &  0.00
\\
	      chain 3 & . & . &0.11 &  0.08 &  0.06 &  0.04 &
 0.03 &  0.02 &  0.02 & 0.01 &  0.01 & 0.01
 &
0.00 &  0.00
\\
	      chain 4 & . & . & . &  0.10 &  0.08 & 0.06 &
 0.04 &  0.03 &  0.02 &  0.02 &  0.01 &  0.01 &
0.01 &  0.00
\\
	      chain 5 & . & . & . & . & 0.10 &  0.08 &
0.05 &  0.04 &  0.03 & 0.02 &  0.02 &  0.01 &
0.01 &  0.01
\\
	      chain 6 & . & . & . & . & . & 0.10  &
0.08 &  0.06 &  0.04 &  0.03 &  0.02 & 0.02 &
0.01 &  0.01
\\
	      chain 7 & . & . & . & . & . & . &
 0.10 & 0.08 &  0.06 &  0.04 &  0.03 &  0.02 &
0.01 &  0.02
\\
 chain 8 & . & . & . & . & . & . &
 &  0.10 &  0.08 &  0.06 & 0.04 &  0.03 &
 0.02 &  0.02
 \\
  chain 9 & . & . & . & . & . & . &
 .&.&  0.10 &  0.08 &  0.06 & 0.04&
  0.03 &  0.02
  \\
   chain 10 & . & . & . & . & . & . &
  . & . &.& 0.10 &  0.08 &  0.06 &
  0.04 &  0.03
\\
   chain 11 & . & . & . & . & . & . &
  .&.&.&.&  0.22 &  0.15 &
     0.06 &  0.04
   \\
   chain 12 & . & . & . & . & . & . &
  . &.&.&.&. &  0.11 &
     0.08 &  0.05
   \\
    chain 13 & . & . & . & . & . & . & .
    &. &. &. &. &. &
      0.10 &  0.07
    \\	
  chain 14 & . & . & . & . & . & . & .
&. &. &. &. &. &.
     &  0.10
   \\
 \hline
	      \end{tabular}
	      \end{footnotesize}
	      \caption{ABC-PT for the modified toy example: the diagonal gives local acceptance rates, and outside the diagonal are given the proportions of accepted exchange moves.}\label{matrixABCPT}
	      \end{table}


\section{Formula for the Tuberculosis example}\label{formuleTuberculose}
	      We used the same notations than \cite{Tanaka2006}. The number of cases of genotype $i$ at time $t$ is denoted by  $X_i(t)$, $G(t)$ is the number of distinct genotypes that have existed in the population up to and including time $t$, and $N(t)$ is the total number of cases at time $t$.
	      \begin{displaymath}
	      N(t)=\sum_{i=1}^{G(t)} X_i(t).
	      \end{displaymath}
	      The genotypes are labeled $1,2,3,\ldots$ for convenience, although the ordering has no meaning, except that $i=1$ represents the parental type from which others are descended (directly or indirectly). The three rates of the system are the birth rate per case per year $\alpha$, the death rate per case per year $\delta$, and the mutation rate per case per year $\theta$. \cite{Tanaka2006} define the following probabilities:
	     \begin{displaymath}
	      P_{i,x}(t) = P(X_i(t)=x),
 \end{displaymath}
	       \begin{displaymath}
\bar{P}_n(t) = P(N(t)=n) {\rm \ and \ }     \tilde{P}_g(t) = P(G(t)=g).
	      \end{displaymath}
	      The time evolution of $P_{i,x}(t)$ is described by the following differential equations:
	      \begin{eqnarray}	
	      \frac{d P_{i,x}(t)}{dt} &=& \underbrace{-(\alpha+\delta+\theta)xP_{i,x}(t)}_\text{no event} + \underbrace{\alpha(x-1)P_{i,x-1}(t)}_\text{birth} \nonumber \\ &&+ \underbrace{(\delta+\theta)(x+1)P_{i,x+1}(t)}_\text{death or mutation},~~ x=1,2,\ldots\\
	      \frac{d P_{i,0}(t)}{dt} &=& (\delta+\theta)P_{i,1}(t).
	      \end{eqnarray}
	      Initially there is only one copy of the ancestral genotype, hence the initial conditions are: $P_{i,x}(0)=0$ for all $(i,x)$, except $P_{1,1}(0)=1$, and for $i=2,3,4,\ldots$, $P_{i,0}(0)=1$.\\
	      To take into account the creation of new genotypes, the probability $\tilde{P}_g(t)$ is described by the following differential equations (only a mutation can create a new genotype):
	      \begin{eqnarray}	
	      \frac{d \tilde{P}_g(t)}{dt} &=& \underbrace{-\theta N(t)\tilde{P}_g(t)}_\text{no mutation} + \underbrace{\theta N(t)\tilde{P}_{g-1}(t)}_\text{mutation}, ~~ g=2,3,4,\ldots\\
	      \frac{d \tilde{P}_1(t)}{dt} &=& -\theta N(t) \tilde{P}_1(t).
	      \end{eqnarray}
	      The initial condition is $G(0)=1$. Let $t_g$ be the time when a new genotype $g$ is created, we have $P_{g,1}(t_g)=P(X_g(t_g)=1)=1$, and $P_{g,x}(t_g)=P(X_g(t_g)=x)=0$ for $x \neq 1$. \\
	      The total number of cases $N(t)$ is described by the following differential equations (only a birth or a death influence changes in this number):
	      \begin{eqnarray}	
	      \frac{d \bar{P}_n(t)}{dt} &=& \underbrace{-(\alpha+\delta)n\bar{P}_n(t)}_\text{no birth, no death} + \underbrace{\alpha(n-1)\bar{P}_{n-1}(t)}_\text{birth} \\
		\nonumber & &  + \underbrace{\delta(n+1)\bar{P}_{n+1}(t)}_\text{death}, ~~ n=1,2,\ldots\\
	      \frac{d \bar{P}_0(t)}{dt} &=& \delta \bar{P}_1(t).
	      \end{eqnarray}
	      The initial conditions are $\bar{P}_1(0)=1$ and $\bar{P}_n(0)=0$ for $n \neq 1$.

\end{document}